\documentclass{jpsj-suppl}
\usepackage{txfonts} 
\usepackage{bm}

\newcommand{\PDG}{Agashe:2014kda}

\usepackage[normalem]{ulem}  
\usepackage[dvips]{color} 

\renewcommand\sout{\bgroup \color{red} \ULdepth=-.5ex \ULset}

\title{Compositeness of Hadron Resonances and Quasi-Bound States}

\author{Tetsuo \textsc{Hyodo}$^{1}$}

\inst{$^{1}$Yukawa Institute for Theoretical Physics, Kyoto University, Kyoto 606-8502, Japan}

\email{hyodo@yukawa.kyoto-u.ac.jp}

\abst{We discuss the composite nature of hadrons appearing near the $s$-wave two-hadron threshold. Generalizing the Weinberg's weak-binding relation for stable bound states, we show that the compositeness of near-threshold resonances and quasi-bound states can be determined by observable quantities.}

\kword{Compositeness, hadronic molecules, low-energy universality, exotic hadrons}

\begin{document}
\maketitle

\section{Introduction}

Stimulated by recent findings of many unconventional hadronic states~\cite{\PDG,Aaij:2015tga}, the structure of excited hadrons draws renewed attention. Of particular interest are the near-threshold states, as represented by the $XYZ$ states~\cite{Brambilla:2010cs}. In the strangeness nuclear physics, the $\Lambda(1405)$ resonance appears near the $\bar{K}N$ threshold~\cite{Hyodo:2011ur}, whose hadronic molecular structure is relevant for the possible existence of the kaonic nuclei~\cite{Akaishi:2002bg,Hyodo:2007jq}.

In recent years, it has been realized that the compositeness of hadrons is useful to unveil the structure of hadrons~\cite{Baru:2003qq,Hyodo:2011qc,Sekihara:2014kya,Guo:2015daa} (see Ref.~\cite{Hyodo:2013nka} for a review). The basic idea goes back to the weak-binding relation for stable bound states derived by Weinberg fifty years ago~\cite{Weinberg:1965zz}. We define the compositeness $X$ as the probability of finding the two-body composite component in the wavefunction of the bound state. It is shown that the compositeness $X$ is related to the scattering length $a_{0}$, the effective range $r_{e}$, and the length scale $R=(2\mu B)^{-1/2}$ associated with the binding energy $B$ and the reduced mass $\mu$ as
\begin{align}
	a_{0} &= R\left\{ \frac{2X}{1+X} + {\mathcal O}\left(\tfrac{R_{\mathrm{typ}}}{R}\right)\right\} , \quad
	r_{e} = R\left\{ \frac{X-1}{X} + {\mathcal O}\left(\tfrac{R_{\mathrm{typ}}}{R}\right)\right\} . \label{eq:Weinberg}
\end{align}
When the binding energy $B$ is so small that $R$ is much larger than the typical length scale of the interaction $R_{\rm typ}$, the structure of the weakly bound state can be determined only from the observable quantities ($a_{0}$, $r_{e}$, $R$) in a model independent manner.

In spite of its potential strength to investigate the hadron structure, the weak-binding relation~\eqref{eq:Weinberg} cannot be directly applied to the unstable states, such as $\Lambda(1405)$. Given the fact that the most of the exotic hadron candidates decay via strong interaction, it is important to extend the applicability of Eq.~\eqref{eq:Weinberg}. Here we present recent studies of the compositeness in a series of works~\cite{Hyodo:2013iga,Hyodo:2014bda,Kamiya:2015aea}, focusing on the generalization of the weak-binding relation to the unstable states.

\section{Near-threshold resonances}

Consider a resonance state close to the lowest energy $s$-wave two-body threshold~\cite{Hyodo:2013iga}. The low-energy two-body scattering amplitude can be well described by the effective range expansion:
\begin{align}
    f(p)
    = & \left[-\frac{1}{a_{0}}-ip+\frac{r_{e}}{2}p^{2}\right]^{-1} ,
    \label{eq:amplitude}
\end{align}
where we adopt the convention of the scattering length as $f(p\to 0)=-a_{0}$. This amplitude has two poles at $p^{\pm} = (i\pm \sqrt{2r_{e}/a_{0}-1})/r_{e}$. When $a_{0}<r_{e}<0$, the pole $p^{-}$ represents the resonance state. 

In contrast to the bound states, the pole position of the resonance state $p^{-}$ determines $a_{0}$ and $r_{e}$ simultaneously. Thus, the compositeness $X$ can be determined only from the pole position:
\begin{align}
    X
    = & \sqrt{
    1-\frac{1}{1-a_{0}/(2r_{e})}}
    =\frac{p^{+}+p^{-}}{p^{+}-p^{-}}
    \label{eq:Xare} .
\end{align}
Unfortunately, however, the compositeness $X$ is obtained as a pure imaginary number. This is a property inherent in the unstable states. On the other hand, from the behavior of the scattering length and effective range of bound states in Eq.~\eqref{eq:Weinberg} with $X\to 1$ (composite dominance) and $X\to 0$ (other cases), we find
\begin{align}
    \begin{cases}
    a_{0}\sim R
    \gg r_{e}\sim R_{\rm typ} & \text{(composite dominance)}\;   \\
    a_{0}\sim R_{\rm typ}
    \ll -r_{e} & \text{(other cases)}\; 
    \end{cases}\label{eq:criterion} .
\end{align}
An important observation is that the scattering length $a_{0}$ and the effective range $r_{e}$ are \textit{real} for the near-threshold resonance state, although the compositeness $X$ and the eigenmomentum $p^{\pm}$ are complex. This means that, instead of examining the complex compositeness $X$, the relation~\eqref{eq:criterion} can be used to judge the structure of near-threshold resonances. The composite dominance is characterized by the small effective range parameter. If the effective range is negative and its magnitude is large, we can interpret that the resonance is not dominated by the two-body composite component. 

As an example, we consider the $\Lambda_{c}(2595)$ resonance in the $\pi\Sigma_{c}$ scattering. Using the central values given by Particle Data Group~\cite{\PDG}, we determine the excitation energy $E=0.67$ MeV and the decay width $\Gamma=2.59$ MeV. Although the compositeness is pure imaginary, $X=0.608i$, we obtain the effective range $r_{e}=-19.5$ fm whose magnitude is much larger than the typical length scale of the strong interaction. Based on the criterion~\eqref{eq:criterion}, we conclude that the $\Lambda_{c}(2595)$ is not dominated by the $\pi\Sigma_{c}$ molecular component.

\section{Near-threshold scaling and zero-energy resonance}

The transition from the bound state solution~\cite{Weinberg:1965zz} to the resonance solution~\cite{Hyodo:2013iga} is studied in Ref.~\cite{Hyodo:2014bda}. Consider that the potential is fine tuned so as to support a bound state exactly at $E=0$, which is called the zero-energy resonance. We then add a small perturbation to the potential $\delta M$. It is shown that the leading contribution to the eigenenergy of the perturbed system $E_{h}$ is expressed by the compositeness of the zero-energy resonance $X(0)$ as
\begin{align}
    E_{h}
    =[1-X(0)]\delta M .
    \label{eq:Zrelation}
\end{align}
Namely, the structure of the zero-energy resonance is reflected in the response of the eigenenergy to the perturbation $\delta M$.

Interestingly, the $s$-wave zero-energy resonance has a special property. The ``compositeness theorem'' proved in Ref.~\cite{Hyodo:2014bda} guarantees that 
\begin{align}
    X(0)
    =1
\end{align}
for a general $s$-wave zero-energy resonance. Intuitively, this is a consequence of the largely extended wave function of a shallow $s$-wave bound state. In the zero energy limit, the radius of the bound state diverges and the composite wave function outspreads to infinity. Any finite contribution of the non-composite component is overwhelmed by the infinitely large composite fraction, and $X(0)=1$ follows. We note that the theorem is valid only for $s$ wave; no constraint on $X(0)$ is imposed in higher partial waves.

Thus, we find the following near-threshold scaling law:
\begin{align}
    E_{h}
    &\propto
    \begin{cases}
     \delta M^{2} & l=0 \\
     \delta M & l\neq 0
    \end{cases} 
    \label{eq:bindingenergy} .
\end{align}
The $s$-wave scaling is qualitatively different from the higher partial waves. This scaling law can also be derived from the expansion of the Jost function. The $s$-wave scaling $E_{h}\propto \delta M^{2}$ indicates that the derivative of the eigenenergy with respect to the perturbation vanishes at origin, $dE_{h}/d\delta M|_{\delta M\to 0}=0$. This forbids the $s$-wave bound state to become directly a resonance state. In fact, the $s$-wave bound state turns into a virtual state, expressed by a pole in the second Riemann sheet of the complex energy plane below the threshold (see also Ref.~\cite{Hanhart:2014ssa}).

\section{Near-threshold quasi-bound states}

So far we have discussed the state around the lowest energy threshold, where the scattering length is real. In a realistic situation, however, we may be interested in a state which lies slightly below the second lowest threshold, but decays into the lowest energy channel. These are called  quasi-bound states. In this case, the scattering length of the closest channel becomes complex.

The generalization of Eq.~\eqref{eq:Weinberg} to the quasi-bound states is given in Ref.~\cite{Kamiya:2015aea} using the effective field theory. The resulting formula for the compositeness $X$ reads
\begin{align}
a_0
&=  R \Biggl\{\frac{2X}{1+X} + {\mathcal O}\left(\left|\tfrac{R_{\mathrm{typ}}}{R}\right| \right) + \sqrt{\frac{\mu^{\prime 3}}{\mu^{3}}} \mathcal{O} \left( \left| \tfrac{l}{R} \right|^{3}\right) \Biggr\}
\label{eq:quasibound} ,
\end{align}
where the complex radius $R=(-2\mu E_{QB})^{-1/2}$ is determined by the eigenenergy of the quasi-bound state $E_{QB}$. The new length scale $l$ is given by the energy difference from the lower threshold $\nu$ as $l=(2\mu \nu)^{-1/2}$. If the absolute value of the eigenenergy of the quasi-bound state is so small that the magnitude $|R|$ is much larger than the interaction range $R_{\rm typ}$ and the scale $l$, we can neglect the correction terms and determine $X$ from the observable quantities, the scattering length $a_{0}$ and the eigenenergy $E_{QB}$.

The compositeness of unstable states should be interpreted with care. As in the case of resonances, the compositeness of quasi-bound state is given by a complex number. With the field renormalization constant $Z$, the compositeness $X$ satisfies the sum rule~\cite{Sekihara:2014kya}
\begin{align}
   Z&+X =1,\quad Z,X \in \mathbb{C} .
   \label{eq:sumrule}
\end{align}
For stable states, $Z$ is interpreted as the probability of finding the components other than the two-body composite. Although the sum of $Z$ and $X$ is still normalized to unity, complex $X$ and $Z$ cannot be interpreted as probabilities. A proposal in Ref.~\cite{Kamiya:2015aea} is to define the real quantities $\tilde{X}$ and $\tilde{Z}$ together with the uncertainty $U$ as
\begin{align}
    \tilde{Z} \equiv &\frac{1 - |X| + |Z|}{2},
\quad \tilde{X} \equiv \frac{1 - |Z| + |X|}{2},
    \quad 
    U \equiv |Z| +|X| -1  .
    \label{eq:newdef}
\end{align} 
When there is a large cancellation in the sum rule~\eqref{eq:sumrule}, $U$ becomes large. On the other hand, if the sum rule is similar to that of the bound state, $U$ becomes small. Thus, $U$ serves as a measure of uncertainty in the interpretation. Moreover, $\tilde{X}$ and $\tilde{Z}$ satisfies the condition for the probabilistic interpretation:
\begin{align}
\tilde{Z}+\tilde{X} =1,\quad \tilde{Z},\tilde{X} \in [0,1] .
\end{align}
In this way, we can interpret $\tilde{X}$ as a probability of finding composite component for the unstable states, as long as $U$ is small.

The $\bar{K}N$ compositeness of the $\Lambda(1405)$ resonance can be studied by Eq.~\eqref{eq:quasibound}. The resonance appears slightly below the $\bar{K}N$ threshold and decays into the $\pi\Sigma$ channel~\cite{Hyodo:2011ur}. The $\bar{K}N$ scattering length and the eigenenergy are recently determined by the systematic theoretical study in Refs.~\cite{Ikeda:2011pi,Ikeda:2012au} as $a_{0}=1.39-i0.85$ fm and $E_{QB}=-10-i26$ MeV. Estimating $R_{\rm typ}$ and $l$ from the $\rho$ meson exchange and the mass difference from the $\pi\Sigma$ channel, we find that the correction terms in Eq.~\eqref{eq:quasibound} are small. We thus determine the $\bar{K}N$ compositeness of the $\Lambda(1405)$ from $a_{0}$ and $E_{QB}$. The result is $X_{\bar{K}N}=1.2+i0.1$, and the corresponding values in Eq.~\eqref{eq:newdef} are $\tilde{X}_{\bar{K}N}=1.0$ and $U=0.5$. This indicates the $\bar{K}N$ molecular structure of $\Lambda(1405)$.

\section{Summary}

We overview the recent studies of compositeness of near-threshold unstable states. The basic formulae~\eqref{eq:Weinberg} and \eqref{eq:quasibound} show that the structure of near-threshold states is reflected in the observable quantities. This is a consequence of the special nature of the low-energy $s$-wave scattering, as shown by the study of the near-threshold scaling and the zero-energy resonance. 

The interpretation of the compositeness of unstable particle is not straightforward, because the compositeness becomes in general complex. It is necessary either to rely upon alternative criteria, or to define some quantities which deserve the probabilistic interpretation. For future perspective, it is important to precisely determine the threshold parameters for the understanding of the nature of the near-threshold exotic hadrons.

\section*{Acknowledgments}
This work is supported in part by JSPS KAKENHI Grants No. 24740152 and by the Yukawa International Program for Quark-Hadron Sciences (YIPQS).



\end{document}